\documentclass[12pt]{article}
\usepackage{graphicx}

\newcommand\pubnumber{Article 21 in eConf C1304143}
\newcommand\pubdate{\today}

\def\apc{APC Universit\'e Paris Diderot, CNRS/IN2P3, CEA/Irfu, Obs de Paris,\\ Sorbonne Paris Cit\'e,
10, rue Domon et Duquet, 75205, Paris, FRANCE}
\def\stan{W.W. Hansen Experimental Physics Laboratory, KIPAC, Dep of Physics and SLAC National Accelerator Laboratory, Stanford University, Stanford, CA 94305, USA}
\def\oab{INAF--Osservatorio Astronomico di Brera, via Bianchi 46, I-23807 Merate, ITALY}
\def\sissa{SISSA, via Bonomea 265, I-34136 Trieste, ITALY}
\def\infn{INFN, Sezione di Trieste, I-34127 Trieste, ITALY}

\def\Title#1{\begin{center} {\Large #1 } \end{center}}
\def\Author#1{\begin{center}{ \sc #1} \end{center}}
\def\Address#1{\begin{center}{ \it #1} \end{center}}

\newcommand\pubblock{\rightline{\begin{tabular}{l} \pubnumber\\
         \pubdate  \end{tabular}}}
\newenvironment{Abstract}{\begin{quotation}  }{\end{quotation}}
\newenvironment{Presented}{\begin{quotation} \begin{center} 
             PRESENTED AT\end{center}\bigskip 
      \begin{center}\begin{large}}{\end{large}\end{center} \end{quotation}}

\def\beq{\begin{equation}}
\def\eeq#1{\label{#1}\end{equation}}
\def\eeqn{\end{equation}}

\def\beqa{\begin{eqnarray}}
\def\eeqa#1{\label{#1}\end{eqnarray}}
\def\eeqan{\end{eqnarray}}

\let\bar=\overbar

\def\Dslash{\not{\hbox{\kern-4pt $D$}}}
\def\dslash{\not{\hbox{\kern-2pt $\del$}}}

\def\msb{{\bar{\ssstyle M \kern -1pt S}}}

\begin{document}
\begin{titlepage}
\pubblock

\vfill
\Title{A hint to the origin of the extended emission in LAT GRBs: the relation between LAT luminosity and prompt energetics}
\vfill
\Author{Lara Nava}
\Address{\apc}
\Author{G. Vianello, N. Omodei}
\Address{\stan}
\Author{G. Ghisellini, G. Ghirlanda}
\Address{\oab}
\Author{A. Celotti}
\Address{\sissa}
\Author{F. Longo, R. Desiante}
\Address{\infn}
\vfill
\begin{Abstract}
We consider the 0.1--10 GeV rest frame light curves of 10 GRBs detected by LAT and with known redshift. In all cases the emission persists after the prompt has faded away. This extended emission decays in time as a power--law. The decay slope is similar among different bursts, while the normalization spans more than 2 decades. However, when the LAT luminosity is normalized to the 1 keV--10 MeV energetics of the prompt emission $E_{\rm iso}$ all light curves become consistent with having the same normalization, i.e. they cluster. At each given time the ratio between the LAT luminosity and the prompt energetics is narrowly distributed. We argue that this result is expected in the external shock scenario and it strengthens the interpretation of the GeV emission in terms of radiation from external shocks. In this context, we derive limits on the distribution of $\epsilon_e$ (the fraction of the shock energy that goes into electrons) and $\eta$ (the efficiency of the mechanism producing the prompt).
\end{Abstract}
\vfill
\begin{Presented}
Huntsville Gamma-Ray Burst Symposium\\
Nashville, USA, April 15--19, 2013 
\end{Presented}
\vfill
\end{titlepage}
\def\thefootnote{\fnsymbol{footnote}}
\setcounter{footnote}{0}
\section{Introduction}
As of August 2011 the Large Area Telescope (LAT) on board the Fermi telescope has detected 28 Gamma--Ray Bursts (GRBs) above 0.1 GeV. The temporal and spectral properties of these events have been presented in the first Fermi-LAT GRB catalog (Fermi LAT Collaboration, 2013).
One of the common characteristic of these bursts is the presence of an extended emission that decays in time as a power-law with a typical slope $\alpha\sim-1$ or steeper in few cases. In three cases a temporal break from a steeper to a flatter decay rate is observed.

Ghisellini et al. (2010) considered the LAT light curves of the 4 brightest GRBs with measured redshift present in their sample (one of which classified as a short burst) and reported evidence of an interesting behavior. When the LAT luminosity (above 0.1\,GeV) is normalized to the total prompt energy $E_{\rm iso}$ (i.e., the energy emitted during the prompt phase in the 1--10$^4$ keV energy range) all the light curves overlap.
They argued that this behavior is expected in the external shock model and it supports the interpretation of the high--energy ($> 0.1$GeV) emission in terms of afterglow radiation. 

In this paper we consider the light curves of all bursts with measured redshift (10 events) presented in the first Fermi-LAT GRB catalog \cite{catalog} and show that the behavior found by \cite{ghisellini10} (that we refer to as 'clustering') is confirmed. We also show that this result is not recovered if we replace $E_{\rm iso}$ with other quantities of the prompt or LAT emission. This suggests that only $E_{\rm iso}$ can be considered a good and robust proxy of the LAT luminosity. We then discuss the theoretical interpretation and the implications of this result.

\section{Sample description and results}
\begin{figure}
\centering
\includegraphics[scale=0.45]{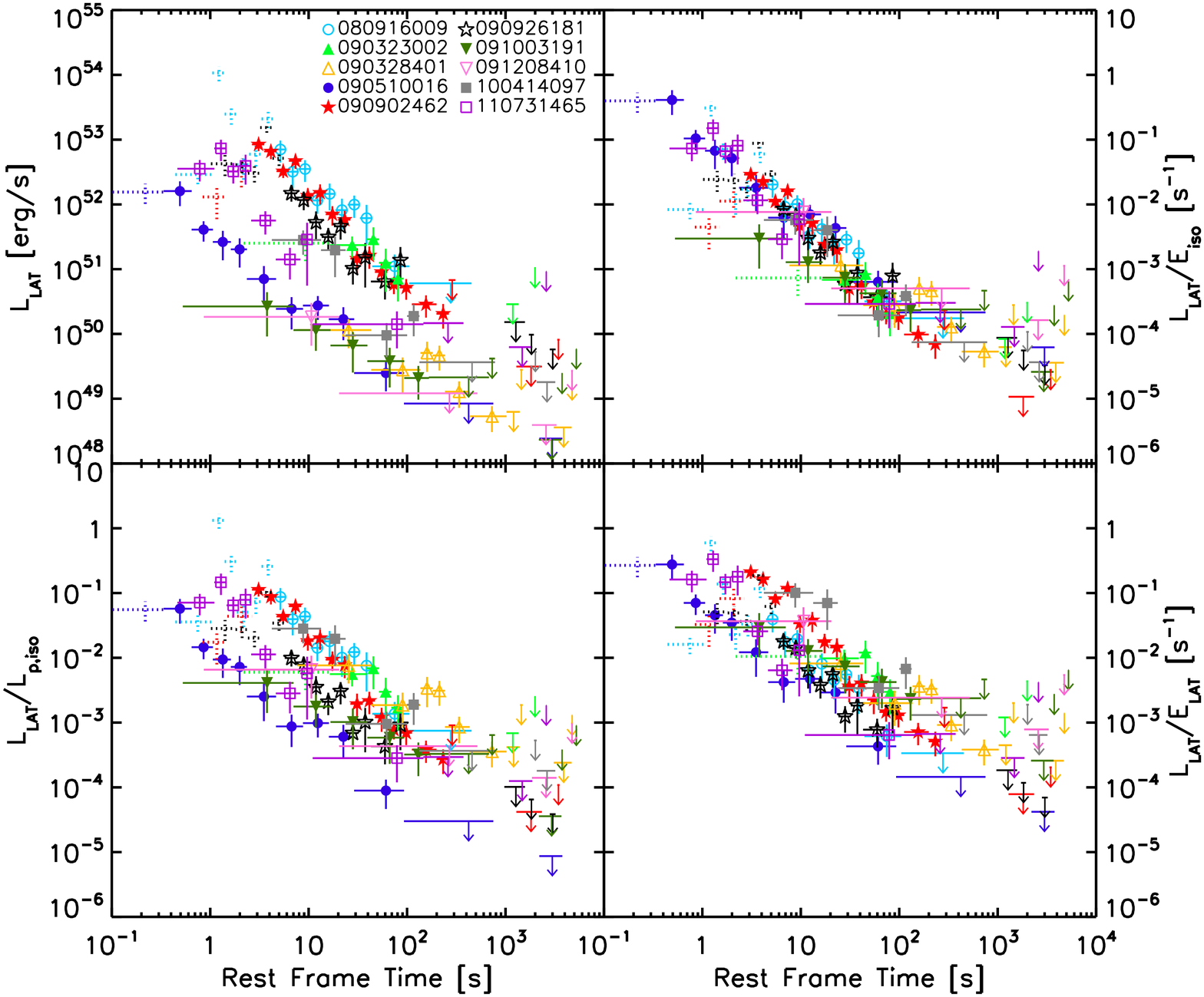}
\caption{Upper left panel: LAT luminosity in the energy range 0.1--10 GeV as a function of the rest frame time for the 10 GRBs with measured redshift present in the first Fermi-LAT catalog. Dotted symbols mark data points not belonging to the power-law extended emission phase. In the other three panels the LAT luminosity is divided by (clockwise): the prompt energy $E_{\rm iso}$, the LAT energy $E_{\rm LAT}$ and the prompt peak luminosity $L_{\rm p,iso}$.}
\label{fig:clustering_four}
\end{figure}

Among all GRBs detected by LAT and analyzed in the first Fermi/LAT GRB catalog \cite{catalog} 10 have measured redshift. We consider the light curves of these 10 bursts (figure~\ref{fig:clustering_four}, upper left panel) and normalize the isotropic LAT luminosity $L_{\rm LAT}$ (estimated in the 0.1--10 GeV energy range) to the total isotropic prompt energy $E_{\rm iso}$ estimated in the 1 keV -- 10 MeV energy range (figure~\ref{fig:clustering_four}, upper right panel). The result found by \cite{ghisellini10} is confirmed: at each given rest frame time the dispersion of $L_{\rm LAT}/E_{\rm iso}$ is significantly smaller than the dispersion of $L_{\rm LAT}$. We also checked if a similar result is obtained when $L_{\rm LAT}$ is normalized to the peak luminosity $L_{\rm p,iso}$ (1 keV -- 10 MeV) of the prompt emission. Only a modest decrease of the dispersion is obtained in this case.

A very simple explanation could be invoked if the LAT emission is the extrapolation at high energies of the prompt spectrum: in this case the fraction of energy falling in the LAT energy range should be nearly the same for all bursts, since they have similar intrinsic properties (they are hard and bright), explaining the common value for the ratio between the emission detected by LAT and by the GBM. However, for 5 of the 10 bursts in our sample the spectral analysis revealed the presence of a spectral extra--component \cite{catalog}, ruling out this explanation.

In general, the luminosity $L_{\rm LAT}(t)$ decays in time as a power--law:
$L_{\rm LAT}(t)\simeq K\,t^{\alpha}$.
The normalization $K$, which spans $\sim$2 orders of magnitude, can be expressed in terms of the total energy $E_{\rm LAT}$ emitted in the 0.1--10 GeV energy range:
\begin{equation}\label{eq:energy}
E_{\rm LAT} \simeq \int_{t_i}^{t_f}L_{\rm LAT}(t)\,dt = K  \int_{t_i}^{t_f}t^{\alpha}\,dt
\end{equation}
Then:
\begin{equation}\label{eq:lum_energy}
L_{\rm LAT}(t)=I(\alpha,t_i,t_f)\, E_{\rm LAT}\, t^{\alpha}
\end{equation}
The value of $I$ (the inverse of the integral in Eq. \ref{eq:energy}, right side) depends on the initial and final times of the LAT emission ($t_i$ and $t_f$) and on the temporal slope $\alpha$. At a given fixed time $t$, the normalization of the curves obtained by dividing $L_{\rm LAT}$ by $E_{\rm LAT}$ is given by $I(\alpha,t_i,t_f)$. Its dispersion depends on the dispersion of $\alpha$, $t_i$ and $t_f$. If for all bursts the LAT emission started and ended at the same $t_i$ and $t_f$ and decayed with the same $\alpha$ then the clustering observed in the $L_{\rm LAT}(t)/E_{\rm LAT}-t$ plane would be perfect. The fact that $L_{\rm LAT}/E_{\rm LAT}$ has a residual dispersion (see the bottom right panel in figure~\ref{fig:clustering_four}) but it is less dispersed than $L_{\rm LAT}$ reflects the fact that different bursts have similar $\alpha$, $t_i$ and $t_f$. One can argue that if there is a strong correlation between $E_{\rm LAT}$ and $E_{\rm iso}$, then one should expect to see a clustering also when $L_{\rm LAT}$ is divided be $E_{\rm iso}$. In this case, the result found by \cite{ghisellini10} and confirmed here would be just a consequence of LAT bursts having similar light curves and of the correlation between the energetics emitted in the two different energy bands. However, by comparing the two right panels in figure~\ref{fig:clustering_four}, it is evident that the clustering in the case of $E_{\rm iso}$ is stronger than the one found when $E_{\rm LAT}$ is considered and it cannot be its consequence. This analysis strongly suggests that at a given rest frame time $L_{\rm LAT}$ is related to the total energy emitted during the prompt. Note that this energy is emitted in a different energy range and on a different period of time. 
This peculiar behavior can be easily explained by considering an external shock origin for the high--energy emission.

\section{Interpretation}
\begin{figure}
\centering
{
\includegraphics[scale=0.41]{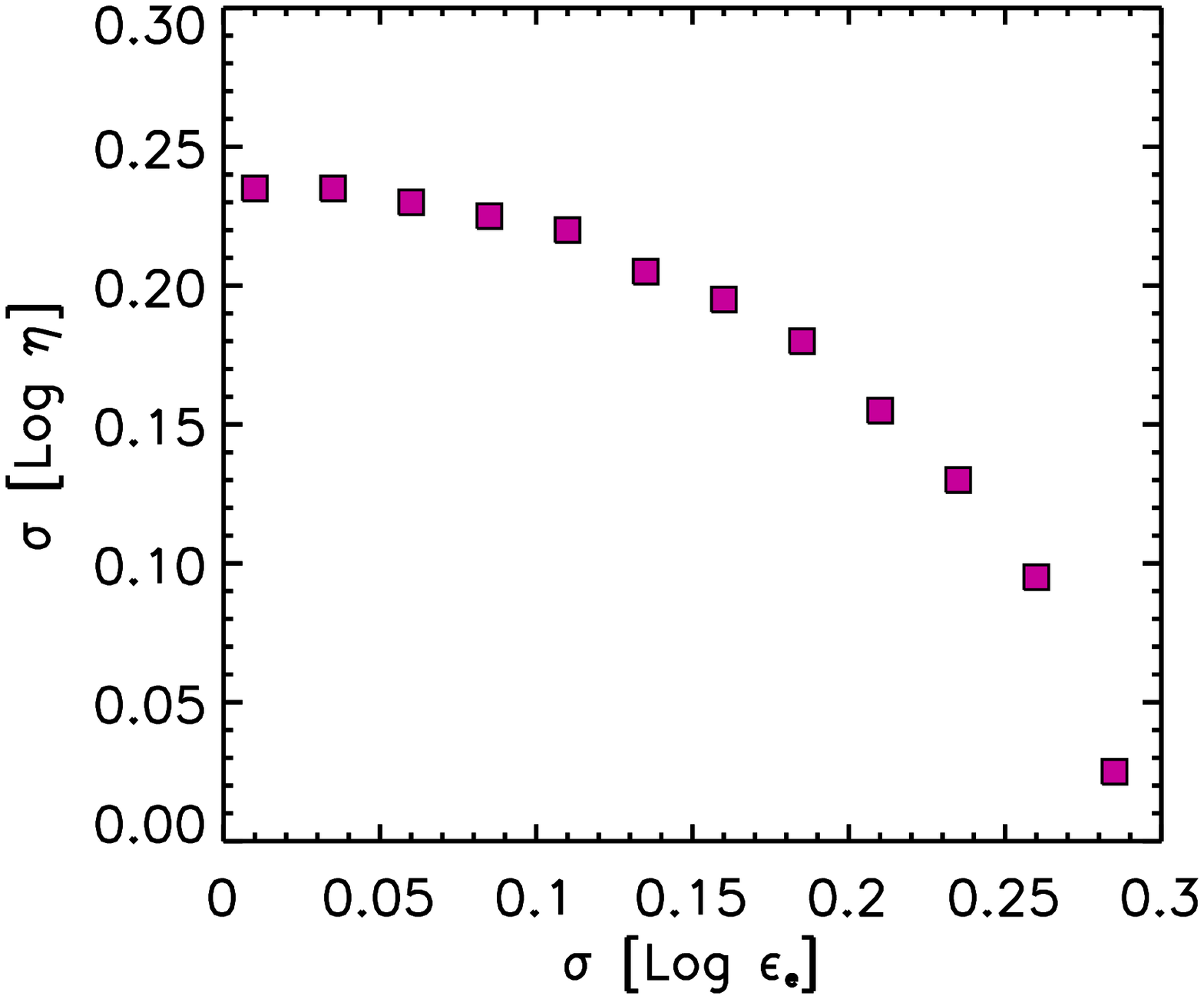}
\includegraphics[scale=0.41]{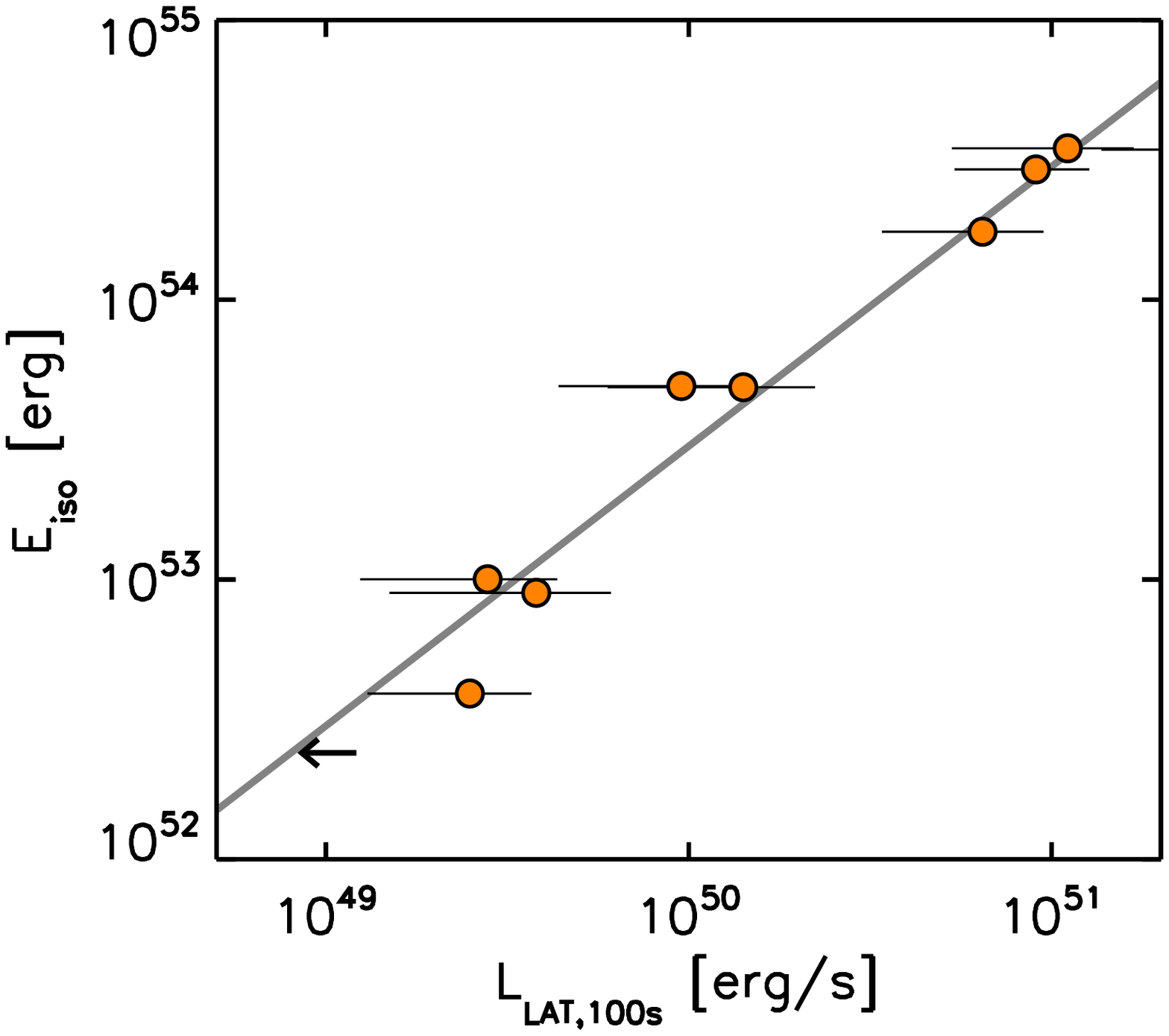}
}
\caption{Left: 1-$\sigma$ width of the distribution of Log$\,\eta$ as a function of the 1-$\sigma$ width of the distribution of Log$\,\epsilon_e$. Right: correlation between the prompt energetics $E_{\rm iso}$ and the LAT luminosity $L_{\rm LAT}$ estimated at the rest frame time $t=100\,$s.}
\label{fig:dispersion}
\end{figure}

In this work we limit our discussion to the scenario in which the LAT emission is synchrotron radiation from electrons accelerated in relativistic shocks driven by an adiabatic blast wave into a homogeneous density medium. A more general discussion will be presented in Nava et al., (in preparation)\index{in prep}. The typical spectral index derived by fitting the Fermi-LAT data with a single power-law model for the bursts in our sample ranges from $\Gamma=-2.2$ to $\Gamma=-2$ \cite{catalog}, suggesting that the LAT luminosity is a good proxy for the bolometric luminosity or that the LAT energy range lies above the characteristic synchrotron frequencies. 
In both cases the light curve of the afterglow radiation is expected to be proportional to $\epsilon_e$ (the fraction of dissipated energy that goes to accelerate the electrons) and $E_{\rm K}$ (the energy content of the blast wave). Since $E_{\rm K}=E_{\rm iso}(1-\eta)/\eta$, the standard afterglow model predicts:
\begin{equation}\label{eq:ratio_th}
\frac{L_{\rm LAT}(t)}{E_{\rm iso}}\propto \epsilon_e\frac{1-\eta}{\eta}t^{-1}
\end{equation}
Therefore, in this model the dispersion on $L_{\rm LAT}(t)/E_{\rm iso}$ is due to the dispersion of the distributions of $\epsilon_e$ and $\eta$. The clustering found in the data suggests that these two parameters must be narrowly distributed. To quantify the maximum width of their distribution we first estimate the vertical scattering of the correlation shown in the right upper panel of figure~\ref{fig:clustering_four}. To this aim, we fit the correlation with a power--law, considering only the data points belonging to the extended emission (solid symbols) and model the vertical scattering of the correlation with a gaussian.
We derive:
\begin{equation}\label{eq:ratio_obs}
\frac{L_{\rm LAT}(t)}{E_{\rm iso}}=0.09~t^{-1.2}  
\end{equation}
The 1-$\sigma$ vertical dispersion is $\sigma$=0.28 and, in the model we are investigating, it limits the width of the distributions of $\epsilon_e$ and $\eta$. Since the individual contribution of each parameter to the total vertical scattering cannot be inferred, we consider a gaussian distribution of Log$\,\epsilon_e$ with $\sigma$ varying from $\sigma_{\epsilon_e}$=0 to $\sigma_{\epsilon_e}$=0.28 (which are the two limiting cases in which the dispersion is entirely due to $\eta$ and to $\epsilon_e$ respectively) and for each value of $\sigma_{\epsilon_e}$ we derive the corresponding maximum value of $\sigma_{\eta}$ from the request that the distribution of Log$[\epsilon_e (1-\eta)/\eta]$ must have a standard deviation $\sigma=0.28$. The result is shown in figure~\ref{fig:dispersion} (left).

\section{Conclusions}
The fact that the afterglow luminosity at frequencies above the characteristic synchrotron frequencies is a good proxy for the energy content $E_{\rm K}$ of the blast wave is a well known result. Estimates of $E_{\rm K}$ are usually inferred from the X-ray luminosity $L_{\rm X}$ at late times (10 hours or later). At these times, in fact, the characteristic frequencies of the synchrotron spectrum generally lie below the X-ray band and this makes $L_{\rm X}$ independent on the external density and weakly dependent on $\epsilon_B$.
This behavior is often presented in terms of a linear correlation between $E_{\rm iso}$ and the $L_{\rm X}$ estimated at a fixed time and it has been used to claim that the efficiency $\eta$ of the mechanism producing the prompt emission must be narrowly distributed \cite{kumar,berger,davanzo,kaneko}. 

In this paper we report on a similar feature discovered in LAT light curves. The correlation between $E_{\rm iso}$ and $L_{\rm LAT}$ (estimated at 100 s) is shown in figure~\ref{fig:dispersion} (right) for the sample of 10 GRBs with measured redshift detected by LAT. The study of this relation allowed us to put constraints on the dispersion of $\epsilon_e$ and $\eta$ (figure~\ref{fig:dispersion}, left). Our analysis is very conservative: since we assume that all the dispersion of the relation is only due to these two parameters.

The LAT light curves have several properties in common with the afterglow light curves, as the smoothness, the presence of an initial peak, the long lasting emission and a temporal decay well approximated by a power--law function. The linear correlation between $L_{\rm LAT}$ and $E_{\rm iso}$ (or, similarly, the clustering of all light curves when the luminosity is divided by $E_{\rm iso}$) is another of these common properties and should be regarded as a further proof that the high--energy emission in GRBs originates from the radiation produced in the external shocks.

\end{document}